# INTEGRATING CAUSAL MACHINE LEARNING INTO CLINICAL DECISION SUPPORT SYSTEMS: INSIGHTS FROM LITERATURE AND PRACTICE

*Completed Research Paper*

Domenique Zipperling, University of Bayreuth and Fraunhofer FIT, Bayreuth, Germany, domenique.zipperling@fit.fraunhofer.de

Lukas Schmidt, Technical University of Munich, Munich, Bavaria, Germany, lukas.s.schmidt@tum.de

Benedikt Hahn, Technical University of Munich, Munich, Bavaria, Germany, benedikt.hahn@tum.de

Niklas Kühl, University of Bayreuth and Fraunhofer FIT, Bayreuth, Germany, kuehl@uni-bayreuth.de

Steven Kimbrough, Wharton School, University of Pennsylvania, Philadelphia, Pennsylvania, United States, kimbrough@wharton.upenn.edu

## Abstract

*Current clinical decision support systems (CDSSs) typically base their predictions on correlation, not causation. In recent years, causal machine learning (ML) has emerged as a promising way to improve decision-making with CDSSs by offering interpretable, treatment-specific reasoning. However, existing research often emphasizes model development rather than designing clinician-facing interfaces. To address this gap, we investigated how CDSSs based on causal ML should be designed to effectively support collaborative clinical decision-making. Using a design science research methodology, we conducted a structured literature review and interviewed experienced physicians. From these, we derived eight empirically grounded design requirements, developed seven design principles, and proposed nine practical design features. Our results establish guidance for designing CDSSs that deliver causal insights, integrate seamlessly into clinical workflows, and support trust, usability, and human-AI collaboration. We also reveal tensions around automation, responsibility, and regulation, highlighting the need for an adaptive certification process for ML-based medical products.*

## 1 Introduction

Artificial intelligence (AI) has increasingly transformed clinical decision-making by enhancing diagnostic accuracy, treatment selection, and patient management (Inoue et al., 2024; Zhang et al., 2024). Building on these advancements, clinical decision support systems (CDSSs) leverage AI to assist clinicians in making evidence-based decisions by integrating medical knowledge and patient-specific data (Elhaddad & Hamam, 2024). However, most existing AI-driven CDSSs rely on associative machine learning (ML) models that focus on pattern recognition rather than causal understanding (Mosqueira-Rey et al., 2023). Although associative ML continues to be used and useful in AI-driven CDSS, these models can only identify correlations and cannot determine causality, making them unsuitable for guiding medical decisions or estimating treatment outcomes (Bica et al., 2020; Sanchez et al., 2022). Consequently, to avoid misleading decision support and its explanations, correlation must be disentangled from causation (Chou et al., 2022). This challenge has motivated recent advances in causal ML, enabling models to reason about cause-effect relationships and quantify treatment effects (Feuerriegel et al., 2024; Sanchez et al., 2022). As a result, causal ML holds promise for improving clinician-AI collaboration (and human-AI collaboration in general) by producing explainable, causally grounded recommendations that foster appropriate and calibrated trust (Amershi et al., 2019; Shin, 2020). Despite this promise, research on causal ML in CDSSs has primarily focused on model development (e.g., algorithms) for specific use cases (Inoue et al., 2024; Zhang et al., 2024) with limited attention to how clinicians can effectively interact with such systems in practice (Feuerriegel et al., 2024; Zheng et al., 2020). For causal ML to enhance clinical decision-making, its predictions and reasoning must be made interpretable and actionable through well-designed interfaces that facilitate trustworthy collaboration (Holzinger et al., 2019; Shin, 2020). To address this gap, our study examines how causal ML can be integrated into CDSS and, in consequence, into medical decision-making by concentrating on human-AI collaboration through the user interface. We, therefore, pose the following research question:

**"How can causal ML-based CDSSs and their explanations be designed to support collaborative decision-making with clinicians?"**

To address the research question, we used a design science research (DSR) approach (Hevner, 2007), combining theoretical insights from existing literature with practical knowledge from the field. We conducted semi-structured interviews with physicians (Whiting, 2008) and performed a structured literature review (SLR), analyzed through open, axial, and selective coding (Gioia, 2021; Strauss & Corbin, 1998; Wolfswinkel et al., 2013), thereby retrieving design requirements (DRs). Finally, we pursued an approach to formulate design principles (DPs) and design features (DFs) grounded in our DRs (Walk et al., 2024). Our analysis revealed three key tensions shaping the integration of causal ML into CDSS: (1) the use-frequency tension, (2) the control-compliance tension, and (3) the generalization-specificity tension. Thereby, we offer a foundation for integrating causal ML into CDSS, highlighting how developers, decision-makers, adjacent managers, and regulatory bodies must adapt to its dynamic, knowledge-driven nature.

# 2 Theoretical Background

CDSSs are software-based tools designed to enhance medical decision-making by providing clinicians with relevant knowledge, patient-specific information, and other health-related data, ultimately aiming to support consistent and optimal patient outcomes (Pierce et al., 2022; Sutton et al., 2020). In general, CDSS are classified into knowledge-based systems using explicit expert knowledge (e.g., using if-then rules) or non-knowledge-based CDSS that infer patterns directly from data. Within this latter category, ML-based CDSS form the predominant subgroup, employing ML algorithms to generate clinical recommendations and support patient care (Berner, 2016). When combined in a collaborative decision-making process, clinicians and ML-based CDSS can outperform both standalone clinicians and purely ML-driven approaches, highlighting the value of complementary human-AI collaboration (Zöller et al., 2025). Realizing this potential, however, requires seamless integration of CDSS into existing medical decision-making workflows (Hemmer et al., 2024), which in turn depends on their acceptance and consistent use by clinicians. Such acceptance largely depends on explainable and interpretable CDSS recommendations, as clinicians must be able to trust the system and justify their decisions to patients, relatives, and stakeholders such as insurance providers (Vellido, 2020).

Most ML-based CDSSs continue to rely on associative ML models that identify statistical relationships between inputs and outcomes, yet do not distinguish whether one variable causes another (Abbas et al., 2025; Holzinger et al., 2019; Sanchez et al., 2022). While post-hoc explanation methods such as feature importance scores and counterfactual examples aim to make model decisions more transparent (Karimi et al., 2021; Mandler & Weigand, 2024), these techniques remain grounded in correlations rather than causality and are associated with major shortcomings (Spitzer et al., 2025). While these methods allow for causal attribution by assigning credit to factors influencing a decision outcome (Weinberg et al., 2025), they cannot ensure that an equivalent real-world intervention would alter a patient's outcome in the same way (Pearl, 2000; Richens et al., 2020). This limitation is especially critical in clinical settings, where causal understanding is essential (Feuerriegel et al., 2024), and correlation-based explanations may risk prompting clinicians to misinterpret associations as causal, fostering illusions of causality (Langer, 1975; Matute et al., 2011; Wang & Mueller, 2016). The emerging concept of causability–the degree to which explanations enable causal understanding for human experts–illustrates this gap: without explicit causal modeling, explanations of associative models may appear plausible yet mislead by implying causation where none has been established (Holzinger et al., 2019). Consequently, the correlation-based foundation of ML-based CDSS constrains their ability to support trustworthy decision-support, thereby motivating growing interest in causal ML, capturing cause-effect relationships more directly and addressing these fundamental shortcomings (Jiao et al., 2024; Weinberg et al., 2025).

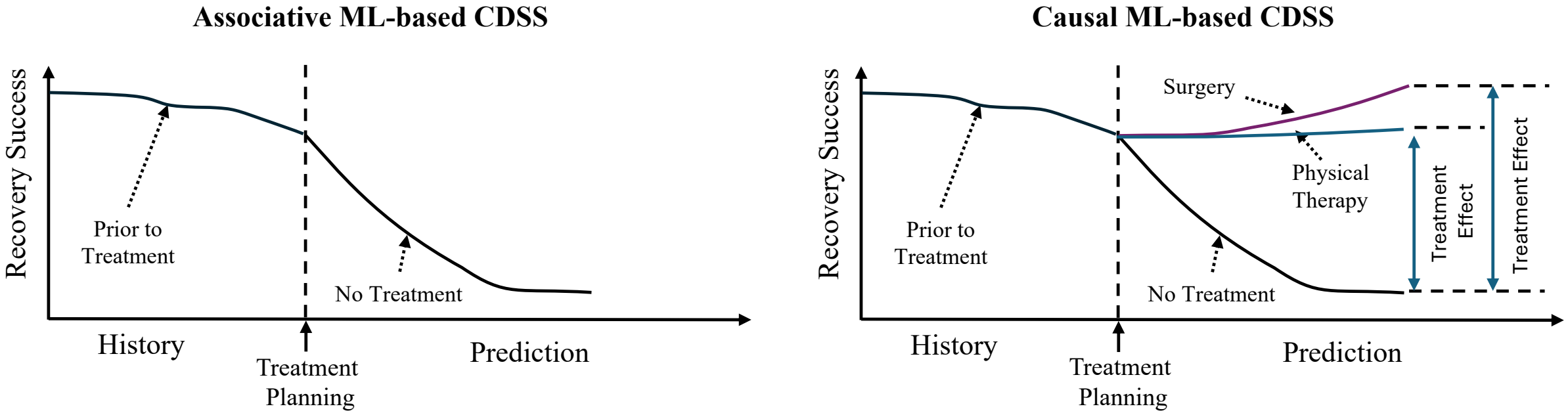

*Figure 1. Visualization of the difference between an associative ML-based CDSS (left) and a causal ML-based CDSS (right) based on Feuerriegel et al. (2024).*

Causal ML extends associative ML by estimating individual treatment effects, enabling clinicians to compare outcomes across treatments and assess the impact of specific interventions (see Figure 1) (Feuerriegel et al., 2024). Yet, realizing this promise remains challenging. Methods such as structural causal models and directed acyclic graphs (DAGs) provide frameworks for representing causal assumptions (Kaddour et al., 2022), but establishing a valid causal structure requires substantial domain expertise to specify, evaluate, and refine these models (Pearl, 2000; Wang & Mueller, 2016). As a result, despite recent methodological advances in causal ML (Inoue et al., 2024; Zhang et al., 2024), these approaches have not been meaningfully integrated into CDSS (Feuerriegel et al., 2024). This gap is not only technical but also socio-technical: if causal ML is to be used in CDSS, clinicians must be able to interact with, understand, and contribute to the causal assumptions embedded in these systems. Explainable AI is important for transparency, actionability, and trust in CDSSs, but does not by itself provide causal reasoning (Petch et al., 2022). Causal ML complements these approaches by making the assumed cause-and-effect relationships underlying model outputs explicit. Because these assumptions are often represented in causal graphs, such graphs are not only a methodological foundation but also a design resource for clinician-facing interfaces. Designing interfaces that exploit this graph structure to support explainable and interactive clinical reasoning, therefore remains an important yet underexplored challenge, motivating our work (Holzinger et al., 2019; Stevens & Stetson, 2023).

# 3 Research Design

We follow Hevner's (2007) DSR guidelines, comprising the relevance, rigor, and design cycles to derive design knowledge on causal ML-based CDSSs (see Figure 2). This study completes the relevance and rigor cycles only to identify the problem domain and derive DRs, preliminary DPs, and potential DFs, thereby providing a base for the subsequent design cycles. To capture clinicians' needs in the relevance cycle, we purposefully sampled physicians developing or testing ML-based CDSS in clinical practice and used snowball sampling after each interview to identify additional relevant participants within their networks (Coyne, 1997). Overall, we interviewed ten physicians for a mean duration of 41 minutes. An overview of the experts can be found in Table 1. Afterward, we anonymized and transcribed all interviews, then coded them using MAXQDA (Kuckartz, 2012).

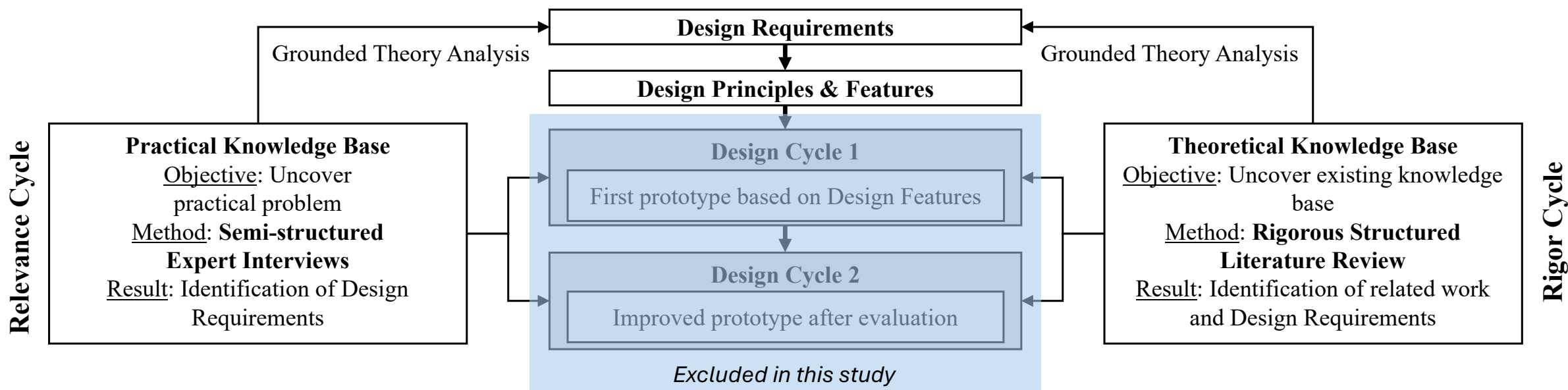

*Figure 2. Methodology based on Hevner (2007). We focus on deriving design requirements, principles, and features rather than conducting design cycles.*

The rigor cycle ensures the theoretical foundation of our work by integrating established scientific knowledge from recent literature. We conducted a rigorous SLR as proposed by Webster & Watson (2002) and Wolfswinkel et al. (2013). We identified the information systems and medical domains as most relevant to our research and selected Scopus, Web of Science, ScienceDirect, and the AIS eLibrary as core and defined a search policy combining concepts of human-AI collaboration, causality, and medicine (see Figure 3). The search was limited to titles, abstracts, and keywords to ensure relevance, except for a full-text search on the AIS eLibrary due to technical constraints. The SEARCH phase yielded 4,321 articles. During the SELECT phase, duplicates were removed and a three-step screening (title, abstract, and full-text) was conducted to retain studies at the intersection of decision support, human interaction, causal information methods, and clinical contexts. Analogous to the relevance cycle, we coded the literature using MAXQDA.

We extracted DRs from expert interviews with physicians and relevant literature using inductive coding, as a deductive approach risked overlooking key requirements (Thomas, 2006), applying the three-step grounded theory approach of open coding, axial, and selective coding (Corley & Gioia, 2011; Gioia, 2021; Gioia et al., 2013; Strauss & Corbin, 1998). The coding results in 343 codes. Focusing on collaborative clinical decision-making, we excluded requirements on causal ML development or IT integration. Axial coding produced 23 second-order themes, aggregated into eight dimensions, interpreted as sub-requirements (SR) and DRs, respectively. Building on Walk et al. (2024), Möller et al. (2020), and (Meth et al., 2015), we follow the general logic of translating DRs into DPs and subsequently into DFs. While Möller et al., (2020) justify the ex-ante formulation of preliminary DPs and DFs prior to artifact instantiation, Meth et al. (2015) inform our use of intermediate translation layers to support this derivation. On this basis, we abstracted the identified DRs to derive preliminary DPs and corresponding DFs for the integration of causal ML into CDSSs.

| Identification | Professional Role | Years of Experience | Duration (min.) |
|---|---|---|---|
| Expert 1 (E1) | Senior Physician | 12 | 52 |
| Expert 2 (E2) | Specialized Physician | 10 | 32 |
| Expert 3 (E3) | Assistant Physician | 5 | 47 |
| Expert 4 (E4) | Specialized Physician | 9 | 40 |
| Expert 5 (E5) | Assistant Physician | 3 | 41 |
| Expert 6 (E6) | Specialized Physician | 8 | 43 |
| Expert 7 (E7) | Assistant Physician | 6 | 36 |
| Expert 8 (E8) | Senior Physician | 22 | 35 |
| Expert 9 (E9) | Assistant Physician | 25 | 35 |
| Expert 10 (E10) | Assistant Physician | 6 | 45 |

*Table 1. Overview of experts and interviews.*

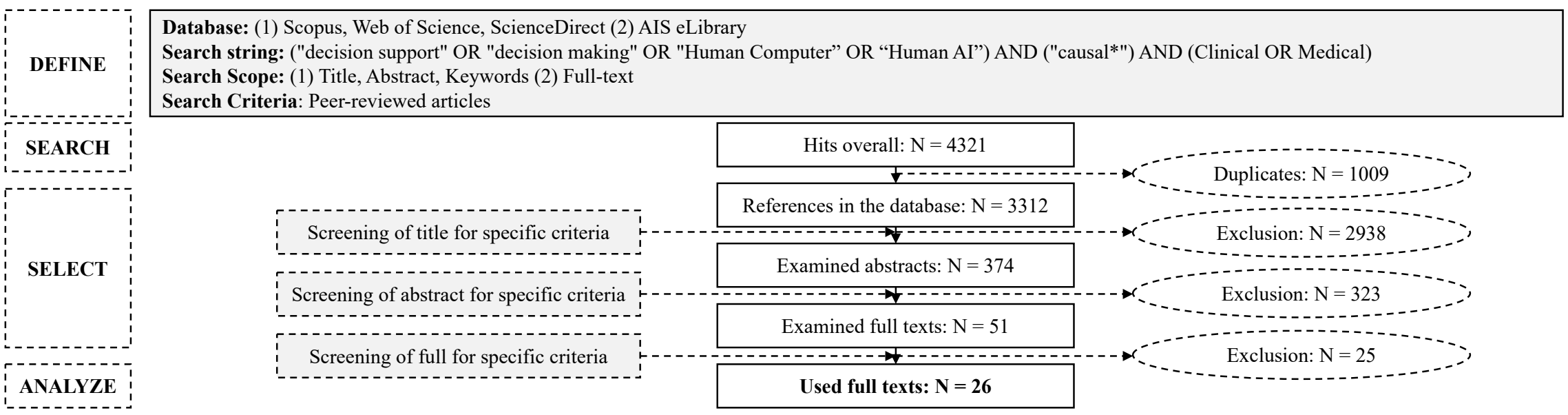

*Figure 3. Rigorous structured literature review based on Wolfswinkel et al. (2013).*

To derive preliminary DPs, we interpreted the eight DRs through five theory-grounded layers: (1) work-system integration, (2) workflow integration, and (3) decision-support transparency, explainability, and clinician trust (Salwei & Carayon, 2022), (4) governance, validation, and monitoring (Labkoff et al., 2024), and (5) causal reasoning added as a causal-ML-specific extension (Sanchez et al., 2022). These layers were used to translate DRs into preliminary DPs and candidate DFs for subsequent design cycles. We understand DRs as class-level artifact requirements that specify what the design should address (Hevner, 2007; Möller et al., 2020), DPs as accessible, prescriptive design knowledge that guides how the artifact should be designed (Gregor et al., 2020; Möller et al., 2020), and DFs as concrete artifact capabilities that operationalize these principles (Meth et al., 2015).

# 4 Results

Here, we illustrate the results by describing the identified DRs, the preliminary DPs, and candidate DFs. Figure 4 provides an overview of the interconnections between DRs, translation layers, DPs, and DFs.

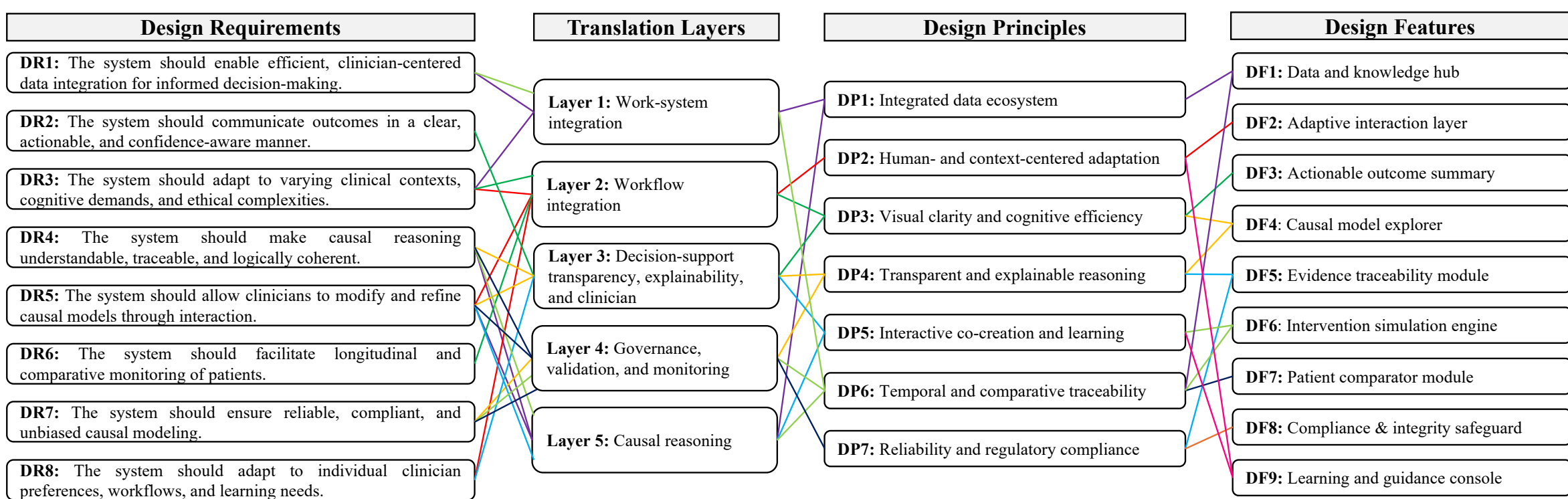

*Figure 4. Overview of DRs, Layers, DPs, and DFs. Colored lines show the specific links.*

## 4.1 Extracted Design Requirements

Within the relevance and rigor cycle, eight DRs were extracted. Table 2 provides an overview of all DRs, their respective SRs, and corresponding sources from literature and expert interviews. Additionally, we group them into three categories: generic, ML-based, and causal ML-based CDSS. Generic DRs mainly address usability, presentation, workflow integration, and interaction. ML-based DRs focus on prediction outputs, explanation, transparency, and governance requirements. Causal ML-based DRs address causal reasoning, intervention effects, and counterfactual exploration.

The first requirement stresses **efficient, human-centered data integration** (DR1), enabling clinicians to combine patient data for decision-making. When automated inputs are incomplete, the system **must support manual updates** (SR1.1), including patient preferences, examination parameters, and clinical measures (E3, E6, Bienefeld et al., 2023), particularly when the algorithm indicates uncertainty or elevated risk (E4). Capturing context and metadata is essential for transparency and understanding decision rationales (Metsch et al., 2024; Müller et al., 2020; Thews et al., 1996). Moreover, **data** **integration must remain human-centered** (SR1.2) by enabling fast, intuitive, and low-effort input, for example, through a few quick clicks, audio input, logic-driven question sequences, graphical relationship visualizations, or natural-language interfaces which can enhance efficiency, transparency, and trust (E5, E6, E10; Vidal et al., 2025; Weiss et al., 1978).

Furthermore, the system should **communicate outcomes clearly, actionably, and with transparency about uncertainty** (DR2). Therefore, the system **must structure and prioritize results** (SR2.1) to enhance interpretability and reduce cognitive load (E1, E4; Pierce et al., 2022; Weiss et al., 1978). For example, key findings should be presented concisely and in a clear order, such as the "degree of the disease, the survival chance, and therapy options" (E9). This can be followed by a summary (E7) or visualization of the "top five features" influencing the decision (E10) in familiar and easily understandable formats, especially for clinicians without ML expertise (Cálem et al., 2024).

| Design Requirement | Sub-Requirement | CDSS Category | Literature | Inter-views |
|---|---|---|---|---|
| **DR1:** The system should enable efficient, clinician-centered data integration for informed decision-making. | **SR1.1:** Supports manual integration and updating of patient-specific data when automated sources are incomplete. | Generic | (Bienefeld et al., 2023; Metsch et al., 2024; Müller et al., 2020; Thews et al., 1996) | 3, 4, 6 |
| | **SR1.2:** Enables fast, natural, and ergonomic data input through intuitive interaction mechanisms. | Generic | (Thews et al., 1996; Vidal et al., 2025; Weiss et al., 1978) | 5, 6, 10 |
| **DR2:** The system should communicate outcomes in a clear, actionable, and confidence-aware manner. | **SR2.1:** Structures and prioritizes results to enhance interpretability and reduce cognitive load. | Generic | (Cálem et al., 2024; Holzinger et al., 2021; Pierce et al., 2022; Thanathornwong, 2018; Weiss et al., 1978) | 1, 3 – 7, 9, 10 |
| | **SR2.2:** Uses multimodal and visual representations to support rapid comprehension and perception. | Generic | (Bienefeld et al., 2023; Pierce et al., 2022) | 1, 6, 7, 9, 10 |
| | **SR2.3:** Translates findings into actionable alerts and recommendations for timely intervention. | Generic | (Bienefeld et al., 2023; Müller et al., 2020; Pierce et al., 2022; Vidal et al., 2025; Weiss et al., 1978) | 1, 3, 5, 7 – 10 |
| | **SR2.4:** Transparently conveys uncertainty and confidence levels to support informed decisions. | ML-based | (Bienefeld et al., 2023; Feuerriegel et al., 2024; Müller et al., 2020; Weiss et al., 1978) | 2, 4, 5, 10 |
| **DR3:** The system should adapt to varying clinical contexts, cognitive demands, and ethical complexities. | **SR3.1:** Dynamically tailores explanations and interactions to the clinical situation and task. | Generic | (Hamon et al., 2022; Pierce et al., 2022) | 1, 3, 4, 6 |
| | **SR3.2** Demonstrates reliability and usefulness to foster clinician trust and adoption. | Generic | (Aslan et al., 2022; Jensen & Andreassen, 2008; Müller-Sielaff et al., 2023; Stevens & Stetson, 2023) | 3, 6, 7, 10 |
| | **SR3.3** Provides reflective and ethical support in complex or uncertain decision contexts. | ML-based | - | 1, 2, 5, 6, 9 |
| **DR4:** The system should make causal reasoning understandable, traceable, and logically coherent. | **SR4.1** Visualizes causal relationships through clear, interpretable diagrams and models. | Causal ML-based | (Bienefeld et al., 2023; Holzinger et al., 2021, 2021; Müller et al., 2020; Müller-Sielaff et al., 2023; Palma et al., 2006; Pfohl et al., 2019; Richens et al., 2020; Sanchez et al., 2022; Vidal et al., 2025; Weiss et al., 1978) | 1 – 7 |
| | **SR4.2** Offers interactive, layered explanations of reasoning processes on demand. | Causal ML-based | (Aslan et al., 2022; Cálem et al., 2024; Hamon et al., 2022; Holzinger et al., 2021; Müller-Sielaff et al., 2023; Palma et al., 2006; Pierce et al., 2022) | 3, 6, 7, 10 |
| | **SR4.3** Links conclusions to verifiable clinical evidence and scientific sources. | ML-based | (Aslan et al., 2022; Müller et al., 2020; Müller-Sielaff et al., 2023; Papageorgiou et al., 2011; Pierce et al., 2022; Vidal et al., 2025) | 1 – 10 |
| | **SR4.4:** Enables interactive exploration of causal networks to support visual reasoning. | Causal ML-based | (Bienefeld et al., 2023; Hamon et al., 2022; Metsch et al., 2024; Müller et al., 2020; Thews et al., 1996) | 1, 10 |
| | **SR4.5:** Maintains logical consistency and causal completeness within reasoning structures. | Causal ML-based | (Palma et al., 2006; Weiss et al., 1978) | - |
| **DR5:** The system should allow clinicians to modify and refine causal models through interaction. | **SR5.1** Supports integration of expert knowledge and evolving medical evidence. | Causal ML-based | (Bienefeld et al., 2023; Cálem et al., 2024; Constantinou et al., 2016; Holzinger et al., 2021; Iakovidis & Papageorgiou, 2011; Metsch et al., 2024; Müller-Sielaff et al., 2023; Palma et al., 2003; Thews et al., 1996; Vidal et al., 2025; Weiss et al., 1978) | 5, 8, 9 |
| | **SR5.2** Enables exploration of causal dynamics through simulations and what-if analyses. | Causal ML-based | (Bienefeld et al., 2023; Cálem et al., 2024; Constantinou et al., 2016; Feuerriegel et al., 2024; Holzinger et al., 2021; Metsch et al., 2024; Müller et al., 2020; Müller-Sielaff et al., 2023; Rabbi et al., 2020; Weiss et al., 1978) | 1, 4, 6, 9, 10 |
| **DR6:** The system should facilitate longitudinal and comparative monitoring of patients. | **SR6.1** Captures and visualizes time-based data to show trends and treatment responses. | Generic | (Bienefeld et al., 2023; Müller et al., 2020; Palma et al., 2003, 2006; Rabbi et al., 2020; Thews et al., 1996; Weiss et al., 1978) | 5, 7 |
| | **SR6.2** Allows comparative evaluation of patient trajectories across cohorts or datasets. | Generic | (Bienefeld et al., 2023; Feuerriegel et al., 2024; Thews et al., 1996) | 10 |
| **DR7:** The system should ensure reliable, compliant, and unbiased causal modeling. | **SR7.1 C**omplies with regulatory and clinical standards to ensure validated performance. | ML-based | - | 3, 4, 8, 9 |
| | **SR.7.2** Prevents bias and data manipulation that could distort causal reasoning. | Causal ML-based | (Dijk et al., 2025; Thews et al., 1996) | 1, 4, 7 |
| **DR8:** The system should adapt to individual clinician preferences, workflows, and learning needs. | **SR 8.1:** Customizes interfaces and explanations to user preferences and roles. | Generic | (Cálem et al., 2024; Jensen & Andreassen, 2008; Vidal et al., 2025) | 3, 9 |
| | **SR8.2:** Integrates smoothly into clinical workflows and task sequences. | Generic | - | 1, 3, 5 – 8, 10 |
| | **SR8.3:** Promotes learning through interactive feedback and explainable guidance. | ML-based | - | 1, 4, 7, 8, 9 |

*Table 2. Overview: Design requirements, their sub-requirement, and the respective source.*

For **multimodal and visual representation** (SR2.2), the system should use intuitive graphics like survival plots, color-coded scales (E7, E9), or simple dashboards (E10). Experts favored "simple, colorful visualizations" that support rapid understanding in time-critical contexts (E1, E6; Bienefeld et al., 2023; Pierce et al., 2022). Further, the system must provide **actionable communication** (SR2.3) by supporting "A/B decisions" (E5, E9), comparing scenarios "with and without surgery" (E10), or trigger alerts when intervention is needed (E7). Finally, **conveying uncertainty and confidence** (SR2.4) is vital for informed decisions. Clinicians requested confidence ranges (E10), probability displays (E4; (Bienefeld et al., 2023; Feuerriegel et al., 2024; Müller et al., 2020), and clear indications of model or data limitations (E4, E5). This fosters transparency, reliability, and accountability, thereby aligning with evidence-based practice in AI-assisted decisions (E2).

Additionally, experts and the literature emphasize the need for a **contextual fit** (DR 3), meaning the system must adapt to varying clinical contexts, cognitive demands, and ethical complexities. For example, **explanations should be context-sensitive** (SR3.1), adapting to different settings ranging from emergencies with minimal input (E1) to large-scale screening (E4), thereby balancing risks and patient preferences (E6). Effective contextualization depends on clarifying the purpose of the explanation (Pierce et al., 2022) and enabling interactive explanations that reflect situational diversity (Hamon et al., 2022). To foster trust, the system must be useful and reliable in practice (SR3.2). Clinicians viewed AI as a means of cross-validating reasoning, "to check if it would have decided like [them]" (E3), and to challenge assumptions constructively (E6, E10). It should feature an intuitive interface (E7) and prioritize clinical needs over technical complexity (Müller-Sielaff et al., 2023). Finally, the system should offer **reflective and ethical support in complex or uncertain cases** (SR3.3). Clinicians valued AI assistance in "borderline decisions" or "when cognitive limits are reached" (E5), helping to prevent fixation errors (E1) while inspiring clinical judgment (E6, E7). In particular, support for rare or ethically challenging cases (E2, E6, E9) can augment reasoning while preserving clinician responsibility.

Another key consideration is **causal reasoning transparency** (DR 4), ensuring that the system's reasoning is understandable, traceable, and logically coherent. Therefore, the system must **clearly visualize causal relationships** (SR4.1) through intuitive, interpretable structures. Experts emphasized graphical or mixed visual-text formats to "grasp information quickly and easily" (E3), preferring "if-then diagrams" or causal graphs over dense text (E5-E7). Graph-based models (e.g., Bayesian networks) can illustrate "cause-effect chains" and "influence factors", while dynamic cues like motion or color highlight important links (E1, E3, E4; Holzinger et al., 2021; Müller-Sielaff et al., 2023). **On-demand explainability** (SR4.2) should let clinicians explore layered explanations, starting with high-level results and drilling into reasoning details as needed (E7, E10; Cálem et al., 2024; Holzinger et al., 2021). Experts requested "short main results first" with the option to "ask questions" or explore deeper reasoning when AI outputs conflict with intuition (E6, E10). Maintaining **evidence traceability** (SR4.3) is crucial and essential for trust and was mentioned throughout all interviews. Clinicians must be able to "check references" (E9) and verify alignment with clinical guidelines and scientific evidence (E10; (Müller et al., 2020; Vidal et al., 2025). Therefore, explanations should link directly to sources (e.g., via buttons or embedded references) and highlight deviations if they are justified (E5, E6). Systems should also update automatically with the latest literature (E1). Finally, **visual interaction** (SR4.4) and **logical completeness** (SR4.5) enable clinicians to navigate and trust the system's reasoning. Experts valued visual drill-down exploration (E1, E10), while dynamic, zoomable graph views and contextual subnetworks enhance navigation (Bienefeld et al., 2023; Metsch et al., 2024). Logical consistency, such as unidirectional causal flow and coherent pathway continuation, preserves interpretive integrity and supports prognostic reasoning (Palma et al., 2003; Weiss et al., 1978).

To ensure models remain clinically relevant, the system must support **modifiable causal relationships** (DR5), allowing clinicians to refine and update causal models as evidence and expertise evolve. To ensure **robust knowledge integration** (SR5.1), the system must enable adaptive yet reliable management of medical knowledge. It should allow expert-guided adjustments to causal models (e.g., nodes, scores, KPIs) while preventing arbitrary edits (E5, E8, E9; Metsch et al., 2024; Müller et al., 2020b). Dynamic, expert-informed knowledge graphs should evolve with new evidence and capture patient-specific nuances through individualized heuristics (Constantinou et al., 2016; Müller-Sielaff et al., 2023; Thews et al., 1996; Vidal et al., 2025). For **causal exploration** (SR5.2), the system should support interactive "what-if" simulations to deepen understanding (E6, E9, E10). Simulating and visualizing alternative scenarios by adjusting factors, observing effects, and tracing outcomes enhances interpretability and learning (Constantinou et al., 2016; Feuerriegel et al., 2024; Müller et al., 2020).

To inform decisions across time and context, the system must support **longitudinal and comparative patient** monitoring (DR6), allowing clinicians to track trajectories and benchmark similar cases. For **longitudinal tracking** (SR6.1), it should visualize temporal dynamics showing patient histories (E5, E7) through timelines or progress bars. Time-stamped data should aggregate into trends that link interventions and outcomes while including changes such as treatment substitutions (Palma et al., 2003; Thews et al., 1996). Temporal granularity must suit the clinical setting, from minute-level intensive care unit tracking to daily follow-ups (E5; Palma et al., 2006). For **comparative evaluation** (SR6.2), the system should enable cohort-level comparisons viewing a patient's progress "against the average" (E10; Bienefeld et al., 2023) to benchmark outcomes and validate effects (Feuerriegel et al., 2024).

Ensuring **causal interaction integrity** (DR7) is essential so that the system's reasoning remains valid, compliant, and protected from manipulation or bias. To **ensure regulatory compliance** (SR7.1), it must meet medical device standards. Clinicians stressed that the CDSS must not "hallucinate" or "improvise," and that model parameters must remain fixed (E3, E8, E9). These measures enhance trust by requiring transparent validation comparable to diagnostic tests, including sensitivity, specificity, and documented limitations (E4), as well as full regulatory approval supported by prospective verification (E3). To **prevent bias** (SR7.2), the system must guard against distortions in data or reasoning (Dijk et al., 2025). Allowing clinicians to alter causal relations can "decrease trust and introduce bias" (E1, E4). Instead, the CDSS should help clinicians identify inconsistencies (e.g., flagging conflicting data, outdated inputs, or implausible links) without enabling unregulated changes (E7; Thews et al., 1996).

**Personalized clinical interaction** (DR8) is essential for adapting the system to preferences, workflows, and learning needs. For **customization** (SR8.1), it should tailor interfaces, explanations, and outputs to different reasoning styles (E9), with adjustable detail, format, and explanation depth (E3; Cálem et al., 2024; Jensen & Andreassen, 2008; Vidal et al., 2025). For **adaptive workflow support** (SR8.2), it must integrate into routines, "running in parallel" with clinical work and providing automated feedback (E5, E6, E8, E10). It should balance autonomy and assistance, offering guidance in complex or educational cases while remaining unobtrusive in routine practice (E1, E3). For **interactive learning and training** (SR8.3), the system should support knowledge sharing and professional development. Clinicians highlighted its value for "guiding young physicians" (E7) and training specialized teams (E4). Short, clickable explanations of results and pitfalls (E1, E4) can enhance understanding, while "knowledge democratization" across institutions promotes shared expertise and continuous learning (E8).

Another output of the relevance cycle is acceptance criteria assessing whether later artifacts improve the environment. For DR1, DR3, and DR8, the artifact should demonstrate acceptable usability (e.g., SUS ≥ 68), work-system/workflow fit, and clinician acceptance through efficient task performance (Bangor et al., 2008; Ji et al., 2021; Salwei & Carayon, 2022). For DR2 and DR4, it should support clear, actionable, and traceable decision-making through correct interpretation of recommendations and uncertainty, rapid identification of relevant actions, and high explanation quality (e.g., System Causability Scale) (Holzinger et al., 2020). For DR5 and DR6, it should enable causal exploration and longitudinal decision support through successful what-if analyses and correct interpretation of intervention consequences, trajectories, and cohort comparisons (Salwei & Carayon, 2022). For DR7, and partly DR4/DR5, it should be reliable and deployment-ready through documented validation, auditability, subgroup checks, and monitoring for safe, compliant use (Labkoff et al., 2024).

## 4.2 Design Principles

The derived DPs translate the eight underlying requirements (DR1-DR8) into an integrated framework for developing a human-centered, causally transparent, and reliable causal ML-based CDSS. Together, these principles ensure that data, knowledge, and reasoning processes are seamlessly combined while remaining understandable, adaptive, and trustworthy for clinicians in real-world practice. A first principle is **an integrated data ecosystem (DP1)** that unifies automated data import, manual clinician input, and scientific knowledge into one coherent structure. This ensures completeness and contextual accuracy while supporting longitudinal patient tracking and cohort-level comparison. By connecting clinical documentation with evolving medical evidence, the system forms a robust foundation for causal reasoning, traceability, and data-driven decision-making. Equally important is **human- and context-centered adaptation (DP2)**. To align with clinical work, all functions must adapt to the clinician's expertise, workflow, and cognitive load. The system should remain intuitive during routine cases yet expand in depth and complexity when expert reasoning or ethical deliberation is required, augmenting rather than interrupting the clinician's thought process. Ensuring **visual clarity and cognitive efficiency (DP3)** is crucial to make complex causal information immediately interpretable. The system should communicate key relationships, factors, and recommendations through perceptually optimized visualizations such as causal graphs or temporal trends. By minimizing visual noise while allowing drill-down for detail, clinicians can grasp essential insights and act quickly under time pressure. Further, the system must maintain **transparent and explainable reasoning (DP4)**. Clinicians need to understand how conclusions are reached, what data support them, and where uncertainty lies. Linking causal explanations directly to scientific evidence, clinical guidelines, and provenance information builds trust, accountability, and traceability, which are key prerequisites for responsible AI in healthcare. To foster continuous improvement and reflective practice, the system should support **interactive co-creation and learning (DP5)**. Clinicians should be able to experiment with causal models, test "what-if" scenarios, and explore intervention effects. This transforms the system from a static recommendation tool into a learning environment that evolves with expertise and medical knowledge, strengthening intuition and confidence, particularly among less experienced clinicians. Moreover, **temporal and comparative traceability (DP6)** enables clinicians to interpret patient development and treatment outcomes over time and against comparable cases. By combining causal reasoning with longitudinal and cohort-level visualizations, the system reveals therapy effects, deviations from expected progress, and cross-patient patterns, supporting both individual care and evidence-based benchmarking. Finally, clinical trust and adoption depend on **reliability and regulatory compliance (DP7)**. The system must adhere to medical device standards, prevent arbitrary manipulation of model parameters, and undergo peer-reviewed validation to ensure safe, reproducible performance. Transparency in sensitivity, specificity, and known limitations reinforces regulatory accountability and user confidence in AI-assisted decision-making.

The seven DPs support a CDSS that integrates data, aligns with clinical reasoning, communicates causality, and maintains ethical and regulatory integrity, enhancing (not replacing) human judgment.

## 4.3 Design Features

Building on the DPs, a set of candidate DFs specifies how the system operationalizes integrated, explainable, and adaptive causal reasoning in clinical decision-making. A **data and knowledge hub** (DF1) unifies automated clinical inputs and clinician-entered information with contextual and evidence-based data into a coherent representation suitable for causal analysis. An **adaptive interaction layer** (DF2) tailors explanation depth, information density, and interaction complexity to clinician expertise, workflow, and clinical setting, supporting both quick glance use and in-depth exploration. To enhance cognitive efficiency and immediate actionability, an **actionable outcome summary** (DF3) aggregates key causal factors, recommended interventions, predicted outcomes, and associated uncertainty into a prioritized, easy-to-interpret view for rapid decisions. Deeper insight into causal structure is provided by a **causal model explorer** (DF4), which enables clinicians to move between overview and detail, inspect relationships and influence paths, and understand the logic behind system reasoning. An **evidence traceability module** (DF5) links conclusions to data sources, guidelines, studies, model assumptions, and known limitations, keeping recommendations verifiable and their scope transparent. Complementing this, an **intervention simulation engine** (DF6) supports safe "what-if" exploration of alternative interventions or parameter changes, allowing clinicians to compare projected outcomes without affecting real patients. A **patient comparator module** (DF7) visualizes patient trajectories over time and benchmarks them against relevant cohorts, connecting causal reasoning with temporal and cross-patient patterns for monitoring and benchmarking. A **compliance and integrity safeguard** (DF8) embeds auditability, bias, and anomaly checks, and regulatory conformity to maintain reliable, non-manipulated causal models and reasoning processes. Finally, a **learning and guidance console** (DF9) leverages these capabilities for education and reflection, offering contextual explanations, micro-tutorials, and case-based exploration that support ongoing professional learning and shared understanding across clinicians and institutions.

To make the derivation logic more transparent, we use DF4 (causal model explorer) as an illustrative example. The feature can be traced back to interview evidence highlighting the need to inspect and verify the system's causal reasoning in more depth when needed. For instance, one expert emphasized the value of being able to "look deeper" into the model's reasoning when human and machine disagree (E1), which was coded as causal exploration and informed SR5.2 under DR5. Another stressed that recommendations should be linked to checkable references to ensure that "there is no error" (E7), informing SR4.3 under DR4. These requirements were then translated through the layers of decision-support transparency, explainability, and clinician trust and causal reasoning, which made their design implications more explicit by foregrounding the need for causal logic that is visible, inspectable, and traceable. On this basis, DR4 and DR5 informed DP3 and DP4, which were operationalized in DF4, the causal model explorer, allowing clinicians to inspect causal structures, examine variable relationships, and trace influence paths within the model.

# 5 Discussion

This study develops design knowledge for human-centered, causal ML-based clinical decision support systems (CDSSs) using a design science research (DSR) approach with rigor and relevance cycles (Hevner, 2007). Based on ten physician interviews and 26 articles from a structured literature review that were jointly coded (Corley & Gioia, 2011; Gioia, 2021), we derived eight design requirements (DRs), seven preliminary design principles (DPs), and proposed nine design features (DFs) that extend generic CDSS design by foregrounding causal reasoning transparency, intervention-oriented support, traceability, and regulatory integrity. While we also identified DRs that are relevant for ML-based or even generic CDSS, we will focus in the discussion specifically on relevant aspects for causal ML-based CDSS and therefore discuss resulting key tensions between DRs as well as their use-case-specific design implications, before outlining contributions and limitations that open avenues for future research.

## 5.1 Tensions

When analyzing the derived DRs and DPs, three tensions emerge, underlining the challenges when designing causal ML-based CDSS. To make these tensions concrete and their implications actionable, we illustrate how their balance differs between the use of a causal ML-based CDSS for individual patient treatment in hospitals and its use in physician-panel deliberations within regulatory bodies and other institutions for guideline development.

**Use-frequency tension.** Clinicians differed in how often the CDSS should be used. While some conceptualized it as a continuously running assistant for all cases, others saw it as most relevant in uncertain or exceptional situations. This reflects prior findings that CDSS must balance automation with selective human engagement to maintain usability and trust (Jensen & Andreassen, 2008; Stevens & Stetson, 2023). This tension, therefore, points to adaptive system behavior that calibrates support to experience and case complexity (DP2). For example, in individual patient treatment, this may imply lightweight support by default and intensified guidance in uncertain or high-risk cases, whereas in physician-panel guideline work, it may imply selective, case-based use rather than continuous assistance.

**Generalization-specificity tension.** The system must balance methodological generality with clinical specificity, as broadly applicable DPs still require tailoring to domain-specific workflows and data structures. For example, the requirements for oncology differ substantially from those in cardiology or psychiatry, particularly regarding temporal dynamics and causal reasoning complexity. This echoes the challenge of combining standardization with contextual adaptation in medical AI design (Hemmer et al., 2024). Across our two exemplary use cases, this implies a modular interface with specialty-specific variables, views, and explanation templates for individual patient treatment, and a configurable analytical environment in which cohorts, endpoints, assumptions, and evidence filters can be adapted to the clinical question under review for physician-panel guideline work.

**Control-compliance tension.** A key tension arises between control and compliance. While interactive editing of causal graphs can enhance clinicians' understanding, unregulated changes risk bias or regulatory violations, as even small changes to a certified CDSS's causal logic can void its medical device certification (DP7), echoing concerns about liability, automation bias, and accountability (Schneeberger et al., 2020; Topol, 2019). Yet, experts emphasize the need to continuously integrate new medical knowledge and expert insights. To balance these demands, systems should separate validated decision modes from exploratory learning environments (Bienefeld et al., 2023), ensuring transparency (DP5) without compromising safety or compliance. When instantiated in our two use cases, this tension implies different forms of interaction: in individual patient treatment, it calls for enabling inspection and simulation without allowing direct modification of the causal logic, whereas in physician-panel guideline work, it calls for a governed update workflow in which proposed changes can be explored in a sandbox environment, versioned, and reviewed before being transferred into the validated system.

This tension also affects regulatory and certification processes for ML-based CDSS. Under the Regulation 2017/745 of the European Parliament and of the Council on Medical Devices (European Union, 2017), even minor modifications to a system's (causal) structure or algorithm may qualify as a "significant change in design or intended purpose" (Art. 120(3)) or a "modification affecting conformity" (Art. 10(10)), requiring reassessment or full recertification. Such obligations conflict with the need for agile, evidence-based updates, as current certification cycles are lengthy, costly, and limited by scarce notified-body capacity (Wien et al., 2023). Their rigidity also hinders the integration of new scientific insights and expert knowledge, reducing the adaptive potential of causal ML and may even weaken the EU's position as a hub for innovation, research, and health (Wien et al., 2023).

Together, these tensions underline that designing causal ML-based CDSSs is not a purely technical challenge but a socio-technical balancing act requiring the careful alignment of automation and human agency, flexibility and regulatory safety, and universality with contextual clinical fit.

## 5.2 Contributions and Managerial Implications

We advance causal ML-based CDSS research by moving from algorithmic proposals to actionable preliminary design knowledge for human-centered, causally transparent, and clinically reliable systems, extending prior work on causal ML in healthcare and causability in medicine (Abbas et al., 2025; Feuerriegel et al., 2024; Holzinger et al., 2019; Richens et al., 2020; Sanchez et al., 2022; Vellido, 2020).

**Practice-Driven Blind Spots.** Most identified DRs are jointly supported by literature and interviews, but differ in their specificity. Some reflect generic CDSS design, such as data integration, communication, and workflow fit. Others align with ML-based CDSS work on explainability, trust, and uncertainty communication. A smaller set is more specific to causal ML-based CDSSs, especially transparent causal reasoning, intervention exploration, and regulatory integrity (Cálem et al., 2024; Constantinou et al., 2016; Mosqueira-Rey et al., 2023; Müller et al., 2020; Müller-Sielaff et al., 2023; Weiss et al., 1978). At the same time, several requirements emerged only in the interviews: reflective and ethical support in borderline cases (SR3.3), explicit regulatory compliance (SR7.1), and embedded workflow and training functions (SR8.2, SR8.3). While prior work addresses trust, responsibility, and legal constraints more generally (Hamon et al., 2022; Pierce et al., 2022; Schneeberger et al., 2020), clinicians expressed more concrete expectations, especially for ethically charged decisions, compliant system behavior, and training support. These interview-only requirements point to future research on ethical decision support, AI-supported reflection, and embedded training features.

**Implications for developers and managers.** For developers, the DPs offer implementable guidance for operationalizing causal ML in CDSS interfaces. DP1–DP3 primarily refine generic CDSS concerns such as data integration, communication, and workflow fit, DP4 addresses more specific ML-based CDSS concerns around interpretability and evidence traceability, and DP5–DP6 foreground causal ML-based CDSS capabilities such as intervention simulation and temporal/comparative reasoning (Cálem et al., 2024; Constantinou et al., 2016; Dijk et al., 2025; Feuerriegel et al., 2024; Müller et al., 2020; Wang & Mueller, 2016). For practice and management, the prominence of DP2 and DP5 underscores that success depends not only on model performance but also on embedding causal ML-based CDSS into workflows, feedback loops, and training structures that allow clinicians to calibrate, challenge, and appropriate AI support (Bienefeld et al., 2023; Sutton et al., 2020; Topol, 2019; Zöller et al., 2025).

**Implications for policymakers.** The interview-based emphasis on strict yet adaptive compliance (DP5, DP7) highlights the need for approval processes that support controlled updates of causal models while preserving safety and traceability. This is especially relevant under the current EU Medical Device Regulation, which largely targets static products and fits only uneasily with adaptive causal ML-based CDSS (European Union, 2017; Hamon et al., 2022; Holzinger et al., 2019; Schneeberger et al., 2020). Future work should explore options, such as sandboxes or staged recertification, aligned with these requirements and assess how they shape clinician trust, accountability, and patient outcomes.

## 5.3 Limitations

This study offers initial insights into the design of causal ML-based CDSSs but has several limitations. First, the empirical basis relies on a small number of expert interviews conducted exclusively with German physicians, which narrows the diversity of perspectives and constrains the generalizability of the resulting DRs, DPs, and DFs. Second, these design artefacts have not yet been validated through prototyping or real-world deployment, so systematic feedback from actual use contexts is still missing. Third, the work deliberately focuses on interface-level requirements and human-AI collaboration, with technical architecture, interoperability, and broader hospital information system integration remaining out of scope. Finally, the interviews were not tied to a specific clinical domain or use case but aimed at deriving domain-agnostic insights. While this supports transfer across settings, it may limit the specificity and immediate applicability of the design guidance for any single application context. Future research should broaden empirical validation by involving more diverse clinical stakeholders and implementing the proposed DFs in a concrete domain-specific prototype to assess usability, acceptance, and decision-support performance under real-world conditions. In addition, examining integration into existing CDSS infrastructures and systematically addressing regulatory and ethical requirements will be crucial to advance clinically viable, trustworthy, and human-centered causal ML-based CDSSs.

# 6 Conclusion

Causal ML can move CDSS beyond opaque prediction toward explicit cause-and-effect reasoning, but its value depends on interfaces clinicians can understand, trust, and integrate into practice. The proposed DRs, DPs, and DFs outline how causal CDSSs can be made more transparent, interactive, and context-sensitive, supporting interpretation, collaboration, and accountable decisions. At the same time, the work exposes tensions that future research must address, for example, reconciling clinicians' desire to adjust causal models with the need to ensure regulatory compliance. Overall, we offer a transferable basis for human-centered causal ML-based CDSSs that aim to augment rather than replace clinical judgment.